\documentclass[namedreferences]{solarphysics}
\usepackage[hyperref,optionalrh,solaromanenum]{spr-sola-addons} 
\usepackage{graphicx}                    
\usepackage{color}                       
\usepackage{breakurl}                         
\usepackage{caption}

\ifx\urlurl    \undefined
\def\urlurl#1{\href{http://#1}{\textsf{#1}}}\fi

\newcommand\nallauth{6986}
\newcommand\ncrit{2007}
\newcommand\nsolar{1910}

\newcommand\ncoun{55}

\chardef\us=`\_

\begin{document}

\begin{article}

\begin{opening}

\title{HelioIndex: A Directory of Active Researchers in Solar and Heliospheric Physics}

%
\author[addressref={aff1,aff2},corref,email={peter.r.young@nasa.gov}]{\inits{P.R.}\fnm{Peter R.}~\lnm{Young}\orcid{0000-0001-9034-2925}}

\address[id=aff1]{NASA Goddard Space Flight Center, Solar Physics Laboratory, Heliophysics Science Division, Greenbelt, MD 20771, USA}
\address[id=aff2]{Department of Mathematics, Physics and Electrical Engineering, Northumbria University, Newcastle upon Tyne, UK}

\runningauthor{P.R.~Young}
\runningtitle{The Solar Directory}

%
\runningauthor{}
\runningtitle{}

\begin{abstract}
HelioIndex is a directory of authors who are active in solar  and heliospheric physics (SHP). It is available at the webpage \url{HelioIndex.org}, and it includes several derived products such as publication lists, country and institute data, journal data, and lists of the most cited articles in the field. HelioIndex is built from ORCID identifiers and publication data obtained from the Astrophysics Data System and ORCID.
Selection criteria have been chosen to approximately correspond to a researcher having completed a PhD and published original research in a refereed journal. HelioIndex is intended to be a comprehensive directory of SHP authors that is generated and maintained through software procedures, with minimal human intervention.  At the time of writing, \nsolar\ SHP researchers are listed in HelioIndex and they belong to  \ncoun\ countries. The countries with the largest numbers of researchers are the US, China and the UK, with 29\%, 15\,\%, and 8\,\%\ of the total, respectively. HelioIndex authors average 0.69 first author papers per year over their careers, and the median citations for a paper is 15. Based on journal keyword data, it is estimated that 57\,\%\ and 28\,\%\ of HelioIndex authors belong to solar physics and heliospheric physics, respectively, with the remainder overlapping both disciplines.
\end{abstract}

%

\end{opening}

\section{Introduction}

A major activity of a research scientist is to perform original research and describe the results in an article submitted to a refereed science journal. In addition to science information, the article includes a set of metadata such as author names, affiliations, journal keywords, journal name, and publication date. Over a career the metadata from the author's ``paper trail" provides valuable data about their career progression. Collecting the metadata from all of the authors in a scientific field potentially yields valuable demographic information that can be tracked in time. For example, which countries and institutes have the most researchers? what journals are most widely used? how many articles are written in the field?

Prior to the digital age, collecting article metadata from paper journals would have been prohibitively time consuming. Since around the mid-1990's metadata have been available in online databases such as  NASA's Astrophysics Data System (ADS), making it simple to collect an author's data by querying on their name. However, authors' names are often not unique. Their paper trails are then tangled and they have to be manually separated by checking affiliation data, for example. Name queries also fail to identify an author's publications if they change their surnames during their career due to marriage. An important advance came with the introduction of the Open Researcher and Contributor ID (ORCID) identifiers in 2009. 

The ORCID organization (\urlurl{orcid.org})  assigns a unique identifier (stylized as ``iD") to scholars who register on their website. The ORCID iD can then be assigned by the scholar to their products, such as articles and proposals. The iDs have become widely used in the solar and heliospheric physics (SHP) field, and most journals encourage authors to assign their iDs at the time of publication. For articles published without ORCID iDs or before 2009, it is possible for authors to insert these articles into their record at \urlurl{orcid.org}. It is also possible to assign ORCID iDs to articles through \href{https://ui.adsabs.harvard.edu/help/orcid/claiming-papers}{an interface at the ADS website}. 

The present article describes an online directory of SHP researchers -- called HelioIndex -- that has been created from their ORCID iDs and article metadata. HelioIndex is generated through automated procedures and is updated on a bimonthly cadence. It presents a snapshot of SHP: where its scientists work; what sub-fields they work in; how many articles they publish; and which journals they use. Over time, HelioIndex will enable trends in the field to be identified, providing data that may be useful to program managers at funding agencies, review committees and other interested stakeholders.

The value of up-to-date demographics data was highlighted in the US National Academies Decadal Survey of Solar and Space Science report \citep{decadal_survey}, where it was found that there is no regular, systematic collection of data by US organizations. Indirect means such as membership in professional societies (the American Geophysical Union and the American Astronomical Society) and NASA proposal submission data were used but these may have systematic biases. The HelioIndex data described here is collected autonomously at a regular cadence, and the sample is well-defined through software algorithms. HelioIndex does not fully address the demographics needs highlighted in the Decadal Survey report. There is no means to determine the gender or race  of community members from publication data, for example, and HelioIndex only covers SHP rather than the entirety of solar and space science (which in the US approximately aligns with the field of Heliophysics). However, HelioIndex may prove a useful tool for future surveys, not only in the US but in all other countries where SHP is performed. 

The only previous studies of publication statistics for the SHP community were by \citet{2016IAUTA..29..245S} and \citet{2016SoPh..291.1267S}. In each case ADS searches were performed to identify solar physics papers based on keywords in the article abstracts. The period 1911--2015 was studied, and the data enabled the number of papers per year and number of unique authors to be tracked in time. The numbers of papers in selected journals per year, and the distribution of papers over countries for 2015 were presented. 

The latest version of HelioIndex is available at \urlurl{HelioIndex.org}. The present article describes how HelioIndex was created, and presents data from the 22 January 2025 release. 
Section~\ref{sect.auth} presents the method for obtaining potential SHP authors, and Sections~\ref{sect.download} and \ref{sect.filter} describe how the authors' publication data are downloaded and filtered to yield the final set of HelioIndex authors. The contents of the HelioIndex webpages are described in Section~\ref{sect.web}, where data and statistics are presented. In Section~\ref{sect:complete} the completeness of HelioIndex in terms of authors and articles is estimated through comparison with the present author's home institute. Section~\ref{sec:compare} compares the results with other available data, and Section~\ref{sect:summary} summarizes the HelioIndex data and discusses how the data may be used to track the evolution of SHP.

\section{Procedure}\label{sect:proc}

The procedure for creating HelioIndex consists of the following key steps:
\begin{enumerate}
    \item Perform searches of the ADS to identify SHP articles, and extract the ORCID iDs of the articles' authors.
    \item For each ORCID iD, download the publication data from the ORCID website and ADS that is related to the  iD.
    \item From the publication data, identify SHP authors based on publication  and article keyword data.
    \item Construct an online database of  SHP authors that includes affiliation information and publication statistics.
\end{enumerate}
These steps are described in the following four sections. HelioIndex is published as a set of html files that are updated on a bimonthly basis. The data from the 22 January 2025 update are presented here, and the HelioIndex files from this date are  preserved at Zenodo (doi:10.5281/zenodo.14740600).

\section{Initial Author Selection}\label{sect.auth}

The initial selection of potential SHP authors is performed through four sets of queries to the ADS.
\begin{enumerate}
    \item Papers published in the past three years that cite the complete set of \emph{Living Reviews in Solar Physics} (LRSP) articles, and the LRSP articles that have been published in the past three years.
    \item All papers published in \emph{Solar Physics} for the current year and the previous year. 
    \item All papers within the last three years that contain the name of a major solar physics mission or observatory in their abstracts. 
    \item All papers within the last three years that contain at least one of a set of SHP-related keywords in their abstracts. 
\end{enumerate}
For queries (iii) and (iv), Table~\ref{tbl.miss-keyw} gives the mission names and keywords that were in use at the time this article was prepared. The keywords were selected to be mostly unique to SHP, limiting the number of authors that would be returned. 
The mission names and keywords are by no means complete in terms of what could be used, and future updates will likely expand on those given in Table~\ref{tbl.miss-keyw}.

The four queries result in a set of bibcodes, which are then submitted to the ADS to return author and ORCID iD data for all of the papers' authors. The names and iDs are stored in the ORCID masterlist file. The format of the name (e.g., whether the forename or an initial is given) is obtained from the author data of the  first article found by the ADS queries in which the author occurs. Diacritical marks are removed from names in order to simplify string matching in the software. Once an author is entered into the masterlist, they remain there in perpetuity. Each of the four queries  is performed every four months, with queries (i), (ii), (iii) and (iv) performed in January, February, March and April, respectively. Also, the queries are performed twice in the same month (separated by two weeks) in case there are technical problems with the ADS API. At the time of preparing this article there were \nallauth\ authors in the masterlist. The particular queries listed above were found to be sufficient to capture a major portion of the SHP community. The queries may be adjusted in the future or replaced with alternative queries.

\begin{center}
\begin{table}[t]
\caption{Missions and keywords used for the initial author selection.}
\begin{tabular}{lp{10cm}}
\noalign{\hrule}
\noalign{\smallskip}
Missions & Hinode, Solar Dynamics Observatory, Interface Region Imaging Spectrograph, Solar Terrestrial Relations Observatory, Solar and Heliospheric Observatory, Parker Solar Probe, Solar Orbiter, RHESSI, Goode Solar Telescope, Inouye Solar Telescope, Wind spacecraft, Advanced Composition Explorer, DSCOVR \\
Keywords & Sunquake, coronal rain, pseudostreamer, Ellerman bomb, supergranule, solar chromosphere, corotating interaction region, ICME, solar flare, Parker spiral, inner heliosphere, heliospheric current sheet, solar energetic particle, coronal mass ejection, strahl \\
\noalign{\hrule}
\end{tabular}
\label{tbl.miss-keyw}
\end{table}
\end{center}

\section{Downloading Publication Data}\label{sect.download}

The first step in downloading publication data for a HelioIndex author is to obtain  a list of the bibcodes of the author's articles. One list is obtained from the \urlurl{orcid.org} API by performing a query on the author's ORCID iD and requesting the list of ``works". Each work may have an external ID, which can be a bibcode or DOI or some other identifier (e.g., an arXiv number). If a work does not have a bibcode, but does have a DOI then the ADS is queried in order to convert the DOI to a bibcode (if available). 
Not all of the entries in the author's ORCID ``works" list will result in a bibcode.
A second list of bibcodes is obtained by querying the ADS API using the author's ORCID iD.
The ORCID and ADS bibcode lists are then combined and duplicate bibcodes are removed. 

The ADS is then queried with the list of the author's bibcodes to yield detailed publication data for each article, including titles, author lists, journal names and citation counts. The ADS field \textsf{doctype} flags the type of document that the bibcode corresponds to, and only the types ``article," ``inproceedings," ``inbook," ``eprint" and ``book" are retained for HelioIndex. Further filtering of the document list is performed to remove certain types of article. For example, articles for which the title begins with ``preface" are removed since they are short, non-science articles introducing a special journal issue or conference proceedings. Articles with ``corrigendum" in the title are removed unless they have been cited. Abstracts submitted to the Solar, Heliospheric, and Interplanetary Environment (SHINE) meetings are incorrectly flagged as ``inproceedings" by ADS and these are also removed.

The publication data for the authors in the masterlist are continually updated through daily calls to the ADS API. Due to limits on the number of queries that can be made to the  API each day, the data for any one author are updated on a 180 day cadence. This means that it will take up to 180 days for a newly-qualified author to transition onto the HelioIndex author list. For authors who are already on the HelioIndex author list (Section~\ref{sect.filter}), however, the update cadence is 45 days in order to keep their publication lists up-to-date.

\section{Filtering the Author Selection}\label{sect.filter}

The masterlist is filtered in two steps to yield the HelioIndex author list. Unlike the masterlist, the author list is generated anew for each HelioIndex update, therefore an author may leave the HelioIndex author list  if they no longer meet the filter criteria. The first filter is based on publication data, and the second is based on journal keyword data. These steps are described in the following subsections.
The publication data and keyword criteria result in \ncrit\ and 3105 of the masterlist authors from making it to the HelioIndex author list at the time this article was prepared. Four additional masterlist entries were excluded because the authors had used two ORCID iDs in their careers, and one of these had been deprecated on the ORCID website. These cases were manually identified and the ORCID iDs were placed in a file such that they will be permanently excluded from all future versions of HelioIndex.

\subsection{Publication Data Criteria}\label{sec:pub-crit}

The publication data used for filtering are the numbers of articles, their type (refereed, first-authored or co-authored), and their publication dates. The ADS field \textsf{property} indicates whether an article is refereed or not.
The ADS assigns the field \textsf{pubdate} to all articles that in most cases gives the publication date as year-month. Hence article ages are usually accurate to within  a month. 
An author's \textit{career age} is defined to be the number of years since the publication date of the author's first first-authored refereed (FAR) article, plus six months.
The latter is added to represent preparation time for the article.  Alternative definitions for career age that were considered were based on the publication date of the author's first article of any type, and the publication date of the author's first refereed article (first-author or co-author). The FAR definition was chosen as it represents a significant landmark in an author's career. The publication dates of authors' first FAR papers varies significantly. For some, the first FAR paper may be produced as part of a Master's degree or even an undergraduate degree, while for others it may appear after the completion of the PhD thesis. However, it is more common for it to appear during the PhD. This will be returned to in Section~\ref{sec:stats}. 

The publication data criteria assigned to determine if an author should be included in HelioIndex are:
\begin{enumerate}
    \item The author's most recent refereed paper must be within three years of the current date.
    \item The author's career age must be at least two years.
    \item The author must have at least one FAR paper.
    \item The author must have achieved at least six ``points," where two points are assigned to FAR papers and one point to co-author papers.
\end{enumerate}
Criteria (i) defines whether a researcher is ``currently active." Three years is chosen to prevent short career breaks such as temporary work reassignments, parenting, and sickness leading to an author leaving HelioIndex. However, this period also means that an author will remain on HelioIndex for some time after permanently leaving the field. 
Criteria (ii)--(iv) were chosen to avoid including PhD students in HelioIndex, in the sense that a Solar Physicist is someone who has obtained the qualification of PhD. In addition, criterion (iii) ensures that HelioIndex members have published their own original research work. For example, a support scientist or engineer may be a co-author on several solar physics papers, but may not have an FAR paper.

FAR articles within the author's publication list are identified  by matching the author's ORCID iD against the ORCID iDs stored with the publication data downloaded from the ADS. If the article does not have ORCID iDs associated with it (such as when the article is listed on the author's ORCID webpage but is not linked to ADS), then the first author status is determined by using the \textsf{author\_norm} field from the ADS output. The latter is a standard form of giving an author's name as ``surname, first initial";  for example, ``Smith, J". It is possible that two authors on the same paper could have the same \textsf{author\_norm} and hence the FAR designation could be wrongfully assigned, but this will be rare.

\subsection{Keywords}\label{sec:keywords}

The initial author selection, although focused on SHP articles, results in many authors from related fields being captured. Stellar and exoplanet astrophysicists are returned as they may reference articles in LRSP, for example. These authors are filtered out through journal keywords assigned to the authors' articles. Most of the major journals used by SHP scientists make use of keywords, including \emph{The Astrophysical Journal}, \emph{Astronomy \& Astrophysics}, \emph{Solar Physics}, and \emph{The Journal of Geophysical Research}. 

\begin{figure}[t]
    \centering
    \includegraphics[width=0.8\linewidth]{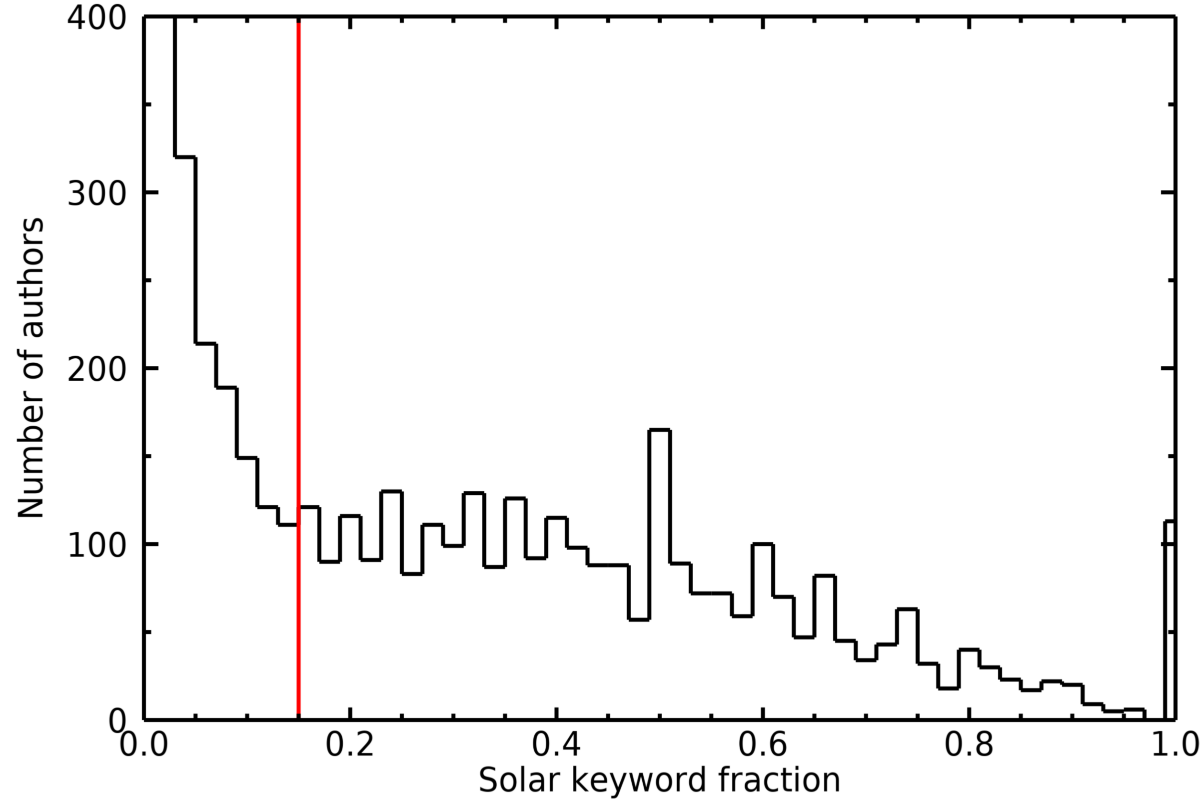}
    \caption{A plot of the solar keyword fractions for the \nallauth\ authors in the masterlist. The plot is truncated in the $y$-direction, with the peak occurring at 2402 authors for a fraction of zero. The red line shows the cutoff value of 0.15 used to identify SHP authors.}
    \label{fig:solar-frac}
\end{figure}

For all of the authors' articles, the keywords are extracted (where available). Numerical keywords, such as from the \href{https://astrothesaurus.org/}{Unified Astronomy Thesaurus}, are removed as the ADS keyword lists include both the numerical and text versions of these keywords. Keywords that contain the strings ``solar and stellar astrophysics", ``solar system" and ``stars: solar-type" are removed as these are often used by stellar Astrophysicists and planetary scientists. 

For the remaining keywords, the fraction of keywords that contain either of ``Sun", ``solar" or ``interplanetary" is computed. If this fraction is greater than 0.15, then the author is considered to belong to the SHP field. 

Figure~\ref{fig:solar-frac} shows the distribution of the solar keyword fraction for all \nallauth\ authors of the masterlist. The cutoff fraction for HelioIndex has been set to 0.15 as this is where the distribution begins to rise towards the large number of authors with low fractions. A list of authors  with fractions between 0.11 and 0.15 was created and their publication and keyword data manually inspected. This generally confirmed that, although their work overlapped with SHP, they mostly work in other fields.

\section{The HelioIndex Webpages}\label{sect.web}

The filtering described in Section~\ref{sect.filter} is performed automatically on the 8th and 22nd of each month to select the list of HelioIndex authors from the full set of authors in the masterlist. The present article shows the results from the update performed on 22 January 2025, which yielded \nsolar\ HelioIndex authors from the \nallauth\ authors in the  masterlist. The most recent version of the directory is provided to the public through the portal \urlurl{HelioIndex.org}. The directory html files from 22 January 2025 are available on Zenodo (doi:10.5281/zenodo.14740600).

\subsection{Author Table}\label{sec:table}

The author table is the principal product of HelioIndex and it gives the authors' names, institutes, countries, fields, and keywords. The format of the author's name is taken from the masterlist (Section~\ref{sect.auth}), and the hyperlink associated with the name goes to a list of the author's publications, sorted by year. Additional links within this page go to the publication list sorted by citation counts, and FAR publication lists sorted by year and citation counts. 
The author's institute and country are obtained through string processing of the affiliation string given in the author's  most recent article (Section~\ref{sec:affil}). If the author has multiple affiliations, then the first one listed is used. Each author is assigned to either solar physics (S), heliospheric physics (H), or both (SH) based on the keywords they assign to their articles. 
The keywords column of the table contains the four  most commonly-used keywords in the authors' articles. 
The table also contains links to the author's ORCID page and a page at ADS that lists the articles associated with the author's ORCID iD in the ADS system. 

The assignment of an author to a field (S, H, or SH) is performed by identifying journal keywords that are considered to uniquely correspond to either solar physics or heliospheric physics. At the time of writing there were 37 and 20 such keywords, respectively, for the two fields. An author is assigned to a field based on the balance between solar and heliospheric keywords in the author's keyword list. If $n_\mathrm{s}$ and $n_\mathrm{h}$ are the numbers of solar and heliospheric keywords, respectively, then we define $f_\mathrm{sh}=n_\mathrm{h}/(n_\mathrm{s}+n_\mathrm{h}$). A researcher is assigned to solar physics if $f_\mathrm{sh}\le 0.20$, and to heliospheric physics if $f_\mathrm{sh}\ge 0.50$. Intermediate values imply the author straddles both fields.

\begin{figure}[t]
    \centering
    \includegraphics[width=0.8\linewidth]{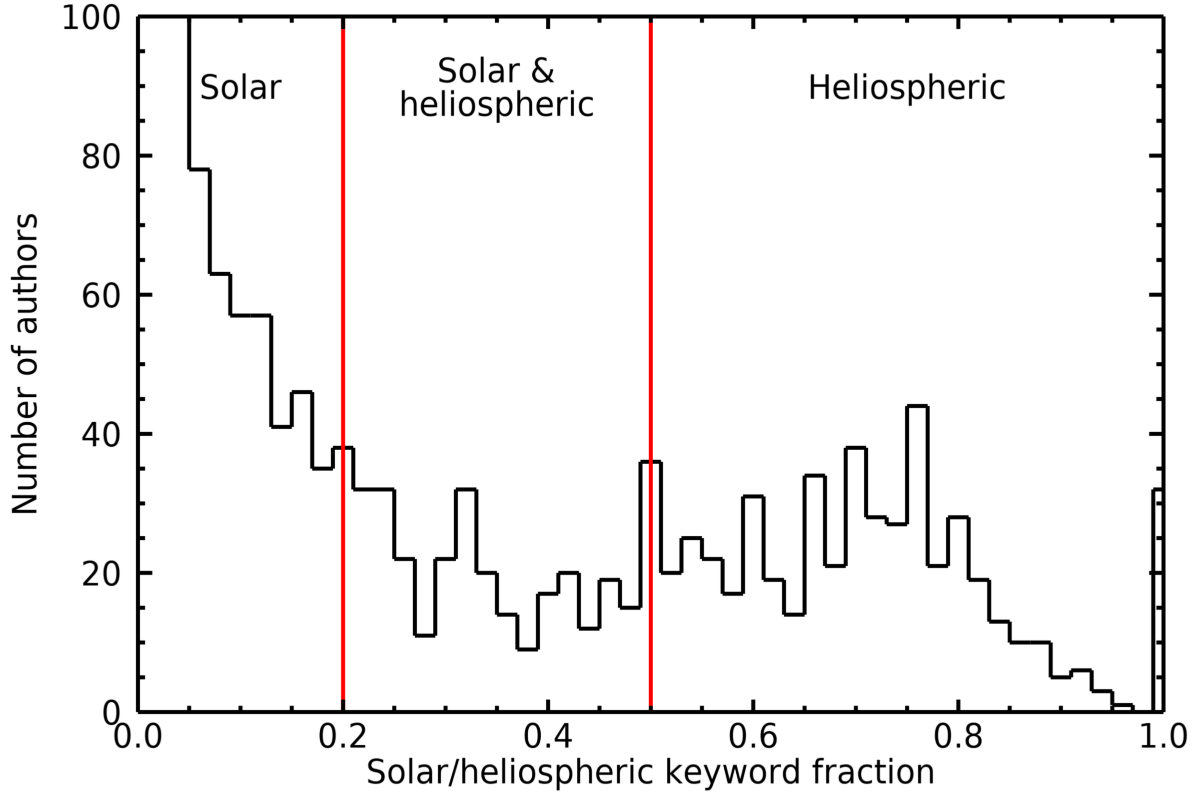}
    \caption{A plot showing the distribution of the solar/heliospheric keyword fraction ($f_\mathrm{sh}$) for the \nsolar\ authors in the HelioIndex author list. The plot is truncated in the $y$-direction, with a peak of 432 for a fraction of zero. The red lines shows the boundaries that separate S, SH and H authors.}
    \label{fig:helio-frac}
\end{figure}

As an example, the present author has a total of 766 keywords, 303 of which are solar keywords and 15 of which are heliospheric keywords. Therefore $f_\mathrm{sh}=0.047$ and an S assignation is made. 
Figure~\ref{fig:helio-frac} shows the distribution of $f_\mathrm{sh}$ values for all HelioIndex authors, with the S and H boundaries indicated. The boundaries were selected by the author based on his knowledge of scientists' expertise and their keyword data.

In the 22 January 2025 release of HelioIndex, 56.7\,\%\ of authors were assigned to solar physics, 27.4\,\%\ to heliospheric physics, and 15.9\,\%\ to both fields. The distributions between the two fields vary with country with, for example, percentages of 47.3/35.1/17.6 (S/H/SH) for the US and 60.1/29.0/11.0 for China, which are the two countries with the largest numbers of solar and heliospheric scientists (Section~\ref{sec:affil}).

\subsection{General Statistics}\label{sec:stats}

The HelioIndex webpage gives a set of key statistics about the authors and their publications, which is reproduced in Table~\ref{tbl.stats}. The statistics focus on FAR papers since each of these is unique in the database: an author's co-authored papers are likely FAR papers of other HelioIndex authors. Statistics are given for how many authors are classed as early-career, mid-career and senior. The Solar Physics Division (SPD) of the American Astronomical Society (AAS) awards prizes for early-career and mid-career scientists and the age ranges are defined relative to the year of award of the scientist's terminal degree. If the latter is awarded in year $Y$, then an early-career scientist's age must be $\le Y+11$, and a mid-career scientist's age must be $\ge Y+12$ and $\le Y+21$. HelioIndex does not have the year of award of degrees, so the year is estimated from the career age (defined in Section~\ref{sec:pub-crit}) and adding 2 years (see below)  to give the estimated PhD date. As an example, the present author's first FAR article was published in July 1997, giving a career start date of January 1997 (Section~\ref{sec:pub-crit}). The estimated PhD date is then January 1999 (the actual award date was October 1998). The early career designation applied until January 2010, and the mid-career designation applied until January 2020. 

The relationship between the career age and PhD award date was investigated by considering the HelioIndex authors who have linked their ORCID iDs to their thesis publications in ADS. There are 194 such authors (only 10\,\%\ of the total). There are some outliers corresponding to senior authors for whom the thesis publication date is much earlier than the first FAR article date, which are due to incomplete ORCID records, and these were removed. For the remaining 174 authors, the mean difference between the thesis publication date and the first FAR article publication date is 2.1~years. In the HelioIndex software the PhD award date is set to the career age plus two years in order to determine the career stage of the author.

\begin{center}
\begin{table}[t]
\caption{HelioIndex statistics for 22 January 2025.}
\begin{tabular}{lc}
\noalign{\hrule}
\noalign{\smallskip}
Statistic & Value \\
\noalign{\hrule}
\noalign{\smallskip}
No. of authors	 & \nsolar\  \\
No. of FAR papers	& 17,605 \\
No. of FAR citations	& 533,494 \\
Average citations per FAR paper	&30 \\
Median citations per FAR paper	& 15 \\
Mean FAR papers per year	& 0.69 \\
Median career age 	& 9.7 \\
Early career researchers	& 62.8\,\% \\
Mid-career researchers	& 20.3\,\% \\
Senior researchers	&16.9\,\% \\
\noalign{\hrule}
\end{tabular}
\label{tbl.stats}
\end{table}
\end{center}

\begin{figure}[t]
    \centering
    \includegraphics[width=0.65\textwidth]{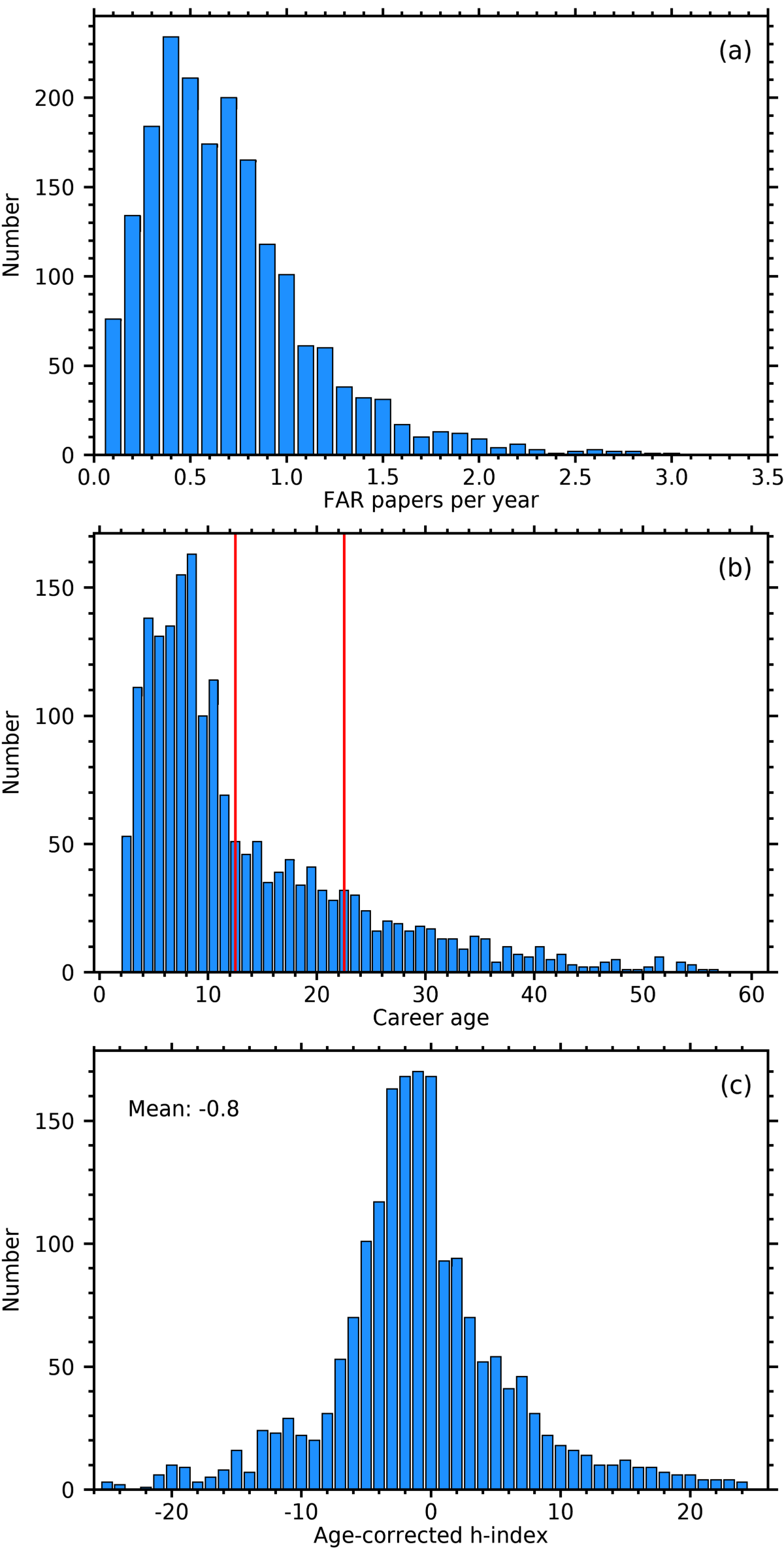}
    \caption{Panel (a) shows the distribution of the FAR papers per year of the HelioIndex authors. Panel (b) shows the distribution of career ages, with the red vertical lines marking the boundaries between early-career, mid-career and senior researchers. Panel (c) shows the distribution of age-corrected $h$-indices for the authors, with the mean value indicated.}
    \label{fig.charts}
\end{figure}

Figure~\ref{fig.charts} shows distributions of three quantities across the HelioIndex members. Panel (a) shows the distribution of FAR papers per year. The mean is 0.69 (Table~\ref{tbl.stats}), implying SHP scientists write, on average, approximately two first-author papers every three years. 18\,\%\ of scientists average more than one FAR paper per year over their careers, and 1.8\,\%\ average more than two FAR papers per year. Figure~\ref{fig.charts} shows the distribution of career ages. The incompleteness of some authors' publication records (Sect.~\ref{sect:complete}) will generally result in their career ages being shorter than their actual career ages, which likely explains the large numbers of researchers around 5 to 10 years. This also means that
the percentages given in Table~\ref{tbl.stats} will be inaccurate. In particular, the number of early-career researchers will be over-estimated and the numbers of mid-career and senior researchers will be underestimated (particularly the latter).

One estimate of a researcher's impact is to compare their $h$-index with their career age. If the $h$-index is larger then it implies the researcher has consistently achieved a number of well-cited articles over their career. In Figure~\ref{fig.charts}c the distribution of the age-subtracted $h$-index is plotted for the HelioIndex authors. The distribution is approximately symmetric, with a mean of $-0.8$. That is, SHP scientists' $h$-indices lag their career ages by 0.8, on average.

\subsection{Country and Institution Statistics}\label{sec:affil}

An author's affiliation is obtained from the \textsf{aff} field of the ADS API output for the author's most recent article. This corresponds to the affiliation assigned to the author in the published article and it usually contains the institute name, the address and the country.  If an author has multiple affiliations, then these are separated by semi-colons in the \textsf{aff} string. For HelioIndex, only the first affiliation is used on the assumption that there is always a primary affiliation that will be listed first by the author.

The HelioIndex software processes the affiliation string to extract the institute name and the country. 
Usually the country can be identified by searching on the country name, but in some cases this does not work. For example, Mexico can be matched with New Mexico State University, which is in the US. In this case, Mexican affiliations are found by matching city or institute names that are unique to Mexico, e.g., Michoacan or Instituto Nacional de Astrof\'isica. The author has created a lookup table of almost 500 entries that enable affiliations to be matched to countries. 

A lookup table has also been created to enable the affiliation string to be matched to an institution name in a standard format. For example, if the affiliation contains ``Goddard" or ``GSFC" then it will be mapped to ``NASA Goddard" in the HelioIndex author table. The institute mapping is performed after the country mapping has been performed. Thus if an affiliation string for an Australian institute contained the word Goddard then it would not be mapped to NASA Goddard. Standardized forms of institution names have not been implemented for all of the HelioIndex author institutes, particularly those with small numbers of scientists. In this case the institute name listed in the author table is obtained by taking the first 40 characters of the affiliation string.

The HelioIndex webpage contains a table that lists all of the countries represented in the directory. For the 22 January 2025 release there are \ncoun\ countries, and those containing  at least 2\,\%\ of the HelioIndex scientists are given in Table~\ref{tbl.coun}. More than half of all researchers are found in just three countries (the US, China and the UK). There are 23 countries with less than five researchers. It is possible that the take up and usage of ORCID iDs varies with country which would lead to varying discrepancies between the true number of SHP researchers and those listed in HelioIndex.

\begin{center}
\begin{table}[t]
\caption{The countries with the largest representations in HelioIndex.}
\begin{tabular}{lrr}
\noalign{\hrule}
\noalign{\smallskip}
Country & Number & \% \\
\noalign{\hrule}
\noalign{\smallskip}
 Total  & 1910 & 100\% \\
US & 558 &    29.2\% \\
China & 283 &    14.8\% \\
UK & 146 &     7.6\% \\
Germany & 96 &     5.0\% \\
Italy & 88 &     4.6\% \\
India & 71 &     3.7\% \\
Spain & 67 &     3.5\% \\
Russia & 62 &     3.2\% \\
France & 54 &     2.8\% \\
Japan & 53 &     2.8\% \\
Belgium & 42 &     2.2\% \\
\noalign{\hrule}
\end{tabular}
\label{tbl.coun}
\end{table}
\end{center}

The institutes with the largest representations in HelioIndex are given in a table linked to the HelioIndex website and a subset is shown in Table~\ref{tbl.inst} for which there are at least 20 staff members, as obtained from the 22 January 2025 release.
Table~\ref{tbl.inst} also shows the percentage of a country's researchers that work in the institute. Germany, Belgium, the Czech Republic and Norway have large fractions of their researchers that work in a single institution. Countries such as the US, China, the UK and Italy are much less concentrated with researchers distributed across many institutes.

\begin{center}
\begin{table}[t]
\caption{Institutes with at least 20 members in HelioIndex.}
\begin{tabular}{llcc}
\noalign{\hrule}
\noalign{\smallskip}
Institute & Country & Number & \%\tabnote{The percentage of authors in the country who belong to this institute.} \\
\noalign{\hrule}
\noalign{\smallskip}
NASA Goddard & US & 64 & 11.5\% \\
MPS - Gottingen & Germany & 47 & 49.0\% \\
Purple Mountain Observatory & China & 34 & 12.0\% \\
National Solar Observatory & US & 30 & 5.4\% \\
KU Leuven & Belgium & 26 & 61.9\% \\
Yunnan Observatories & China & 26 & 9.2\% \\
U. Alabama - Huntsville & US & 25 & 4.5\% \\
University of Oslo & Norway & 24 & 96.0\% \\
National Space Science Center & China & 24 & 8.5\% \\
Harvard-Smithsonian CfA & US & 23 & 4.1\% \\
Southwest Research Institute & US & 23 & 4.1\% \\
IAC - Tenerife & Spain & 23 & 34.3\% \\
SSL, Berkeley & US & 23 & 4.1\% \\
NAOC & China & 22 & 7.8\% \\
NJIT & US & 22 & 3.9\% \\
MSSL & UK & 21 & 14.4\% \\
University of Michigan & US & 20 & 3.6\% \\
Astronomical Institute Ondrejov & Czech Rep. & 20 & 71.4\% \\

&& \\
\noalign{\hrule}
\end{tabular}
\label{tbl.inst}
\end{table}
\end{center}

\subsection{Journal Statistics}\label{sec:journal}

Table~\ref{tbl.jour} gives the most-commonly used journals for the two most recent, complete years. Only FAR articles are included in order to reflect the journals that HelioIndex authors choose to publish in for their own work.  In comparison to the statistics provided by \citet{2016SoPh..291.1267S} using a different method, a key difference is the much lower numbers of articles for journals such as \textit{Journal of Geophysical Research}, \textit{Geophysics and Aeronomy}, and \textit{Geophysical Research Letters} found here. These journals are typically used by Earth scientists, suggesting that the Schrijver statistics may include articles related to magnetospheric and ionospheric physics that are outside of the scope of HelioIndex.

\begin{center}
\begin{table}[t]
\caption{HelioIndex journal statistics.}
\begin{tabular}{lcc}
\noalign{\hrule}
\noalign{\smallskip}
& \multicolumn{2}{c}{Percentage of articles} \\
\cline{2-3}
\noalign{\smallskip}
Journal & 2023 & 2024 \\
\noalign{\hrule}
\noalign{\smallskip}
ApJ & 41.5 & 45.3 \\
A\&A & 14.2 & 16.7 \\
Solar Physics & 5.9 & 7.3 \\
MNRAS& 4.5 & 3.5 \\
\noalign{\smallskip}
No. of articles & 994 & 935 \\
\noalign{\hrule}
\end{tabular}
\label{tbl.jour}
\end{table}
\end{center}

\begin{figure}
    \centering
    \includegraphics[width=0.7\linewidth]{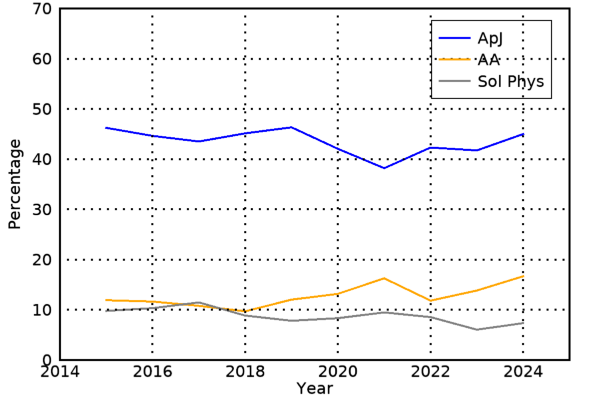}
    \caption{The percentages of articles written by HelioIndex authors and published in three journals as a function of time for the past ten years. The journals are \textit{The Astrophysical Journal} (ApJ; blue), \textit{Astronomy and Astrophysics} (AA; orange) and \textit{Solar Physics} (Sol Phys; grey).}
    \label{fig:jour}
\end{figure}

Three journals are consistently the most used by SHP authors: \textit{The Astrophysical Journal} (ApJ), \textit{Astronomy and Astrophysics} (AA) and \textit{Solar Physics}. Figure~\ref{fig:jour} shows how the percentages of SHP articles written by HelioIndex authors and published in these journals has changed over the past ten years. 

\subsection{Most-cited Articles}

Citation counts for the articles written by HelioIndex authors are downloaded with the ADS data. Tables have been constructed for the HelioIndex webpage showing the twenty most-cited articles for each of the past ten years. In addition there are tables for the most-cited articles for all of the past ten years and all of the past five years. Often the most-cited articles are those describing missions, instrumentation, and software packages. Review articles written for \textit{Living Reviews in Solar Physics} and \textit{Space Science Reviews} also feature prominently.

\subsection{Articles in Science and Nature}

The journals \emph{Nature} and \emph{Science} are the highest impact-factor journals available to SHP authors, and articles in these journals are often considered particularly significant in an author's career. These articles  are identified and compiled into a  table that is available on the HelioIndex website. The list is manually inspected to remove ``comments," ``perspectives," and ``news and views" articles so that only original research articles are retained.
Eighty-one articles are listed, dating back to 1972. HelioIndex authors have averaged 1.8 articles per year in these journals over the past ten years. The median citations of all the articles is 105, significantly above the median citations of all the HelioIndex author articles (Table~\ref{tbl.stats}).

\section{Completeness of the Directory}\label{sect:complete}

The author works at NASA Goddard Space Flight Center (GSFC), which has the largest number of SHP researchers in the world (Table~\ref{tbl.stats}). Based on the author's knowledge of colleagues at GSFC, an estimate of the completeness of HelioIndex can be made. 

The staff working at GSFC consist of government civil servants and contractors employed by different institutions.  At the time of writing, there were 34 civil servant researchers who the author  considered to be active SHP researchers and be expected to be included in HelioIndex. Seven (21\,\%)  of these are currently missing from HelioIndex. One scientist does not have an ORCID iD, and the remaining six have not linked all of their publications to their ORCID iDs. 
It was confirmed that if these authors had linked all of their publications, and the queries described in Section~\ref{sect.auth} had found the authors, then they would have appeared in HelioIndex.

For each of the 27 civil servants who are listed in HelioIndex, the author used ADS to find all of their FAR papers. The total is 528, which compares with 300 papers found by the HelioIndex software, i.e., HelioIndex has found only 57\,\%\ of the FAR papers. One senior author accounts for 27\,\%\ of the FAR paper total, but most of these papers are missing from HelioIndex. Excluding this person, HelioIndex contains 73\,\%\ of the FAR papers. This should be considered  more representative of the completeness of HelioIndex as a whole. The FAR publication list is complete for 12 of 27 civil servants. Generally, it is the younger scientists who have all of their publications linked to their ORCID iDs.

The GSFC civil servants have a median career age of 20, significantly higher than the median career age of the HelioIndex members (Table~\ref{tbl.stats}), which suggests the missing FAR papers for HelioIndex are likely smaller than for the GSFC cohort. Thus the estimate of 27\,\%\ missing FAR papers is likely an upper limit on the actual number of missing papers.

In summary, the GSFC sample suggests that up to 21\,\%\ of SHP researchers are missing from HelioIndex either because they do not have ORCID iDs or because they have not linked all of their articles to their ORCID iDs. Of those authors that are included in HelioIndex, the GSFC data suggest up to 27\,\%\ of their articles may be missing from HelioIndex.

\section{Comparisons with Other Sources}\label{sec:compare}

In this section, comparisons are made with demographic data from other sources. The 2024 US National Academies report \citep{decadal_survey} discussed demographics of the US solar and space science community, and Section~4.2.1 presented estimates of the size of the workforce. A 2011 survey was sent to 2560 unique email addresses, and there were 1171 responses from people based in the US. If SHP is considered to constitute half of solar and space science, then these numbers suggest a US SHP population of between 586 and 1280. The report also gave data from NASA showing the number of unique names that have submitted proposals to the NASA heliophysics division each year from 2011 to 2021. The number varies between 750 and 1000, with 913 for the most recent year. Dividing by two to estimate the SHP component gives 457. The survey and NASA values are thus comparable to the 558 US-based HelioIndex members (Table~\ref{tbl.coun}).

Another source of data is membership of professional societies. The author does not have access to data from the American Geophysical Union, which has a Space Physics and Aeronomy section that includes a Solar and Heliospheric physics subsection, but does have access to data from the AAS/SPD. As of 2025 January, the SPD has 264 full members and 45 affiliate members (i.e., people who are members of the SPD but not the AAS). This number excludes students, emeritus members and international affiliates and thus is most directly comparable to the US-based HelioIndex authors. The list of members was matched against the HelioIndex author list using first and last names, and members based outside of the US were excluded. In total, there are 149 SPD members (52\,\%\ of the US-based SPD members) listed in HelioIndex. A detailed study of the SPD members who are not listed in HelioIndex was not performed, but spot checks on individual members showed a number of stellar astrophysicists and non-scientists in the SPD list, and scientists who did not meet the publication or keyword criteria for HelioIndex. Only 26\,\%\ of US HelioIndex scientists are SPD members, which suggests that the SPD member list alone will not yield accurate demographic data for the US SHP community.

The \citet{2016SoPh..291.1267S} study gave around 4050 SHP authors worldwide for 2015, more than double found here. This difference was not investigated in detail, but the journal statistics presented in Section~\ref{sec:journal} suggest the Schrijver study included papers outside of SHP, for example magnetospheric physics, that resulted in more Geophysics journals than found in HelioIndex. The Schrijver study also includes PhD students that are mostly excluded from HelioIndex (Section~\ref{sec:pub-crit}).

\section{Summary and Discussion}\label{sect:summary}

A directory of currently active researchers in SHP called HelioIndex has been presented. The method for creating the directory using mostly autonomous software algorithms has been described, and some statistical results presented. HelioIndex is available at the website \url{HelioIndex.org} and is updated on a bimonthly cadence.

The accuracy of the HelioIndex data depends on authors linking their publications to their ORCID iDs. Based on the sample of civil servant scientists at NASA GSFC, it is estimated that 21\,\%\ of SHP scientists may be missing from HelioIndex due to incomplete publication records. In addition, up to 27\,\%\ of FAR papers may be missing. If the use of ORCID iDs continues to increase, the accuracy and completeness of HelioIndex data will improve. However, data such as number of FAR papers per year (Table~\ref{tbl.stats}), the relative numbers of solar and heliospheric scientists (Section~\ref{sec:table}), and the lists of countries and institutes with the most researchers (Tables~\ref{tbl.coun} and \ref{tbl.inst}) are unlikely to change significantly.

At present, HelioIndex gives a current snapshot of SHP scientists and their work. A future development will be to present trends in key parameters with time. For example, total numbers of scientists, numbers of scientists in each country, number of publications, median age and median citations. Because the take-up of ORCID iDs is relatively recent it does not make sense to study trends over the past 20 years, for example, as many scientists who were active in this period but have left the field never had an ORCID iD and will be unlikely to create one in the future. Their work will therefore never be visible to the HelioIndex software. As the HelioIndex data accumulates over the coming years, clear trends in the data should become apparent and will be provided on the HelioIndex website. The software has the feature that the various tables can be generated for any user-specified date but using the most recent set of ADS publication data. For example, suppose an author had an incomplete set of publications associated with their ORCID iD on 8 August 2024, but subsequently assigned their iD to all of their publications. When the HelioIndex data was originally generated on 8 August  it would have had access to the author's incomplete publication data. Regenerating the 8 August data today, however, the software would have access to the complete publication data. 

HelioIndex is not intended to be a definitive record of people active in SHP. As highlighted in Section~\ref{sect.filter}, it mostly excludes PhD students and support scientists, as well as citizen scientists who make valuable contributions to the field. In addition, HelioIndex is solely focused on published research articles, but there are many more activities performed by SHP workers that are not captured, including public outreach work, engineering, telescope operations, and teaching. These activities can not be identified or quantified through automatic procedures as done here, but require alternative means such as community surveys. In this case HelioIndex should provide a valuable point-of-reference against which these  data can be compared.

\begin{ack}
I thank the referee and C.\ Paulsen Young for valuable comments. The research has made use of NASA's Astrophysics Data System Bibliographic Services.
\end{ack}

\begin{dataavailability}
\end{dataavailability}

\begin{ethics}
\begin{conflict}
The author declares that he has no conflicts of interest.
\end{conflict}
\end{ethics}

\bibliographystyle{spr-mp-sola}
\bibliography{ms}{}

\end{article} 
\end{document}